# Layered semiconducting electrides in p-block metal oxides

Jiaqi Dai[1,2,], Feng Yang[1,2], Cong Wang[1,2], Fei Pang[1,2], Zhihai Cheng[1,2] and Wei Ji[1,2],*

[1]*Beijing Key Laboratory of Optoelectronic Functional Materials & Micro-Nano Devices, Department of Physics, Renmin University of China, Beijing 100872, China*

[2]*Key Laboratory of Quantum State Construction and Manipulation (Ministry of Education), Renmin University of China, Beijing 100872, China*

*Emails: wji@ruc.edu.cn (W.J.)

**ABSTRACT:**

**In conventional electrides, excess electrons are localized in crystal voids to serve as anions. Most of these electrides are metallic and the metal cations are primarily from the s-block, d-block, or rare-earth elements. Here, we report a class of p-block metal-based electrides found in bilayer SnO and PbO, which are semiconducting and feature electride states in both the valence band (VB) and conduction band (CB), as referred to 2D "bipolar" electrides. These bilayers are hybrid electrides where excess electrons are localized in the interlayer region and hybridize with the orbitals of Sn atoms in the VB, exhibiting strong covalent-like interactions with neighboring metal atoms. Compared to previously studied hybrid electrides, the higher electronegativity of Sn and Pb enhances these covalent-like interactions, leading to largely enhanced semiconducting bandgap of up to 2.5 eV. Moreover, the CBM primarily arises from the overlap between metal states and interstitial charges, denoting a potential electride and forming a free-electron-like (FEL) state with small effective mass. This state offers high carrier mobilities for both electron and hole in bilayer SnO, suggesting its potential as a promising p-type semiconductor material.**

In certain ionic compounds, some electrons delocalize from their parent atoms and occupy interstitial sites within the lattice[1-3]. These excess electrons, confined within cavities (0D)[4-7], channels (1D)[8,9], layers (2D)[10-13] or network (3D)[14], act as anions and interact ionically with surrounding metal cations. These compounds, known as electrides, exhibit unique properties like high conductivity[10,15,16], low work function[12,17,18], and distinctive catalytic activities[19-21], making them attractive for applications in electronics[22], energy storage[23,24], and catalysis[25,26]. Due to weak attractions by the atomic nuclei, these excess electrons, referred to as electride electrons, were often found to occupy energy states around the Fermi level, which results in them exhibiting metallic or semi-metallic properties. The ionic interactions from the cations have minimal impact on the energy levels of the electride electrons, maintaining the electrides (semi-)metallic.

To form an electride, the preferred oxidation states and stoichiometry of the component atoms must provide at least one electride electron per formula unit. The cations are typically electropositive metals from the s-block[27] (e.g., Li, K, Mg, Ca), d-block[28,29] (e.g., Sc, Y, Zr, Hf), or rare-earth elements[30,31] (e.g., Gd, La). Among electrides comprised of these metals, the cation and anion interactions are not solely ionic. Hybrid electrides, such as $Y_2C$[12] and $Hf_2S$[32], exhibit electronic hybridizations between the electronic states of the metal cations and the electride electrons. Stronger hybridization occurs with more electronegative metals[33], such as Sc and Al, which results in the splitting of the (semi)metallic states into bonding and antibonding states, thereby transforming the material into a semiconductor. Notable examples include $Sc_2C$ (bandgap of 0.31 eV) and $Al_2C$ (0.50 eV). The magnitude of the bandgap is directly related to the strength of this hybridization, raising an important question: What is the upper limit of electronegativity for metals that can serve as cations in electrides? Answering this question is critical for determining the maximum bandgaps achievable in electrides, clarifying the limit for potential applications of electrides in future electronics.

P-block metals typically exhibit higher electronegativity than those from the s- and d-block metals. A previous theoretical prediction has considered Al, a p-block metal, identifying its layered carbide, $Al_2C$, as an electride candidate[33]. However, the localization of excess electrons in $Al_2C$ (0.3 for the electron localization function, ELF) is less pronounced compared to conventional electrides (>0.5 for the ELF[34]). This raises an intriguing question of whether p-block metals can serve as effective cations in electrides with electron localization comparable to that of conventional electrides. Here, we report the identification of two layered semiconducting p-block electrides, SnO and PbO, which have been experimentally synthesized. They represent exemplary candidates for this category of materials, and similar structured materials could likely be discovered through high-throughput computational screening, as elucidated elsewhere. Electronic localization function (ELF) maps, plotted at isosurface values of 0.60 and 0.70 for SnO and PbO, respectively, clearly indicate that the electride electrons are confined within the van der Waals gaps between layers in SnO and PbO. Notably, the degree of electron localization in SnO and PbO is comparable to that of conventional electrides and much larger than that of $Al_2C$. Further analysis of the electronic structures reveals that these electride electrons participate in covalent-like interactions with the metal cations. Accordingly, we refer to SnO and PbO as "covalent-like electrides", which distinguishes them from both the traditional ionic bonding observed in metallic electrides and the hybrid ionic bonding found in few known semiconducting electrides. These covalent-like interactions also give rise to free-electron-like interfacial states within the valence and conduction bands.

**Results and Discussion**

Each of SnO and PbO exists in multiple polymorphic forms, with the α phase (litharge) being the most stable under ambient conditions. These two oxides share a layered tetragonal crystal structure (space group P4/nmm, Fig. 1a), where each metal atom is coordinated by four oxygen atoms in a square planar geometry, and the layers adopt an AA stacking sequence. In the case of bulk SnO, the fully relaxed in-plane lattice constant is 3.816 Å, which decreases slightly to 3.796 Å in the bilayer (BL) and to 3.780 Å in the monolayer (ML), reflecting a compression of approximately 0.5 % compared to the bulk phase. The PbO crystal exhibits a similar trend, with its fully relaxed bulk in-plane lattice constant of 4.001 Å reducing to 3.986 Å in the BL and 3.976 Å in the ML, corresponding to a compression of about 0.6 %. Figure 1a (1b) shows the top (side) views of the fully relaxed atomic structure of a SnO (PbO) bilayer, in which the metal atoms of the upper layer are positioned directly above the oxygen atoms of the lower layer, exhibiting an interlayer distance of 2.942 Å (2.618 Å).

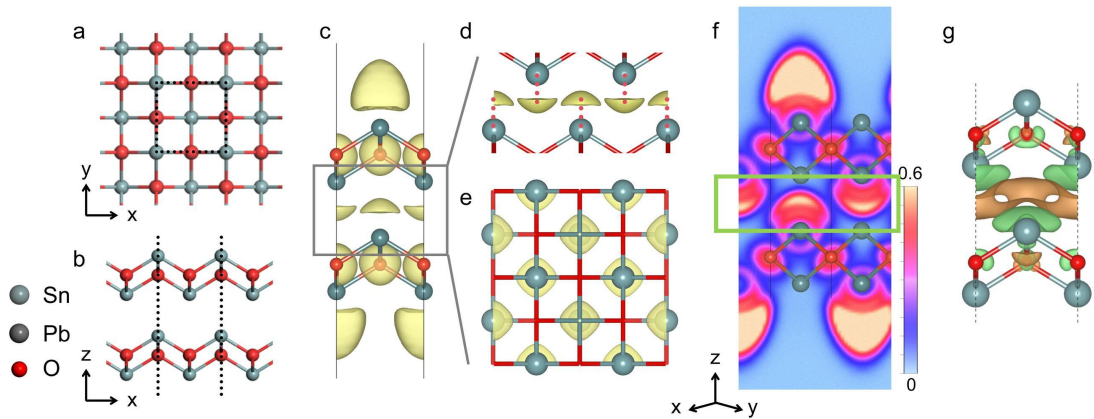

**Figure 1. Atomic structures and ELF for bilayer SnO and PbO. a-b**) Top view (a) and side view (b) of the atomic structures for α-SnO. Sn, Pb, and O atoms are represented by grey, black, and red spheres, respectively. c) 3D maps of the ELF for bilayer SnO. The isosurface contours are 0.6. d-e) Zoom-in of the gary box in (b) from the side view (d) and the front view (e). The interlayer excess charge and the Sn atoms directly aligned with it are marked with red dashed lines. f) 2D maps of ELF for bilayer SnO from another perspective. The green box highlights the excess electrons distribution between layers, located at 2c position, with values of 0.6. g) Interlayer differential charge densities of bilayer SnO. The isosurface contours are $5 \times 10^{-4}$ e/Bohr$^3$.

The ELF is of paramount importance to exclusively identify electrides[34,35]. Figure 1c plots an isosurface contour of the ELF for SnO with an isosurface value of 0.6, explicitly showing the existence of localized electrons within the interlayer region. The interfacial ELF contours appear tetragonally umbrella-like and equally distributed within the interface, as illustrated in Fig. 1c and 1d for side- and top-views. Each of the contours develops from an interfacial Sn atom of one layer toward its superposing O atom in the other layer, occupying the 2c Wyckoff positions, specifically located at the coordinates (1/4, 1/4, z) and (3/4, 3/4, z). This distribution indicates that the interlayer excess electrons predominantly exhibit 2D characteristics, confined to specific regions between the layers, in other words it is a 2D electride candidate.

From an ELF slicing shown in Fig. 1f, the excess electrons of bilayer SnO differ from those in conventional electrides, where excess electrons typically interact ionically with cations and exhibit discontinuity in electron density distributions. In contrast, in bilayer SnO, the excess electron densities overlap with those of the interfacial Sn atoms, suggesting electronic hybridizations between the states of excess electrons and Sn electrons. Figure 1g shows the interfacial differential charge density (DCD) of bilayer SnO, revealing a pronounced covalent-like characteristic in charge redistribution at the vdW interface. This characteristic is marked by localized charge depletion around the interfacial Sn atoms and charge accumulation in the interfacial region. The significant interfacial electron accumulation suggests that the formation of interlayer excess electrons is likely related to the stacking of the two layers, where the electrons tend to localize in the interlayer region. More interestingly, charge accumulation is observed not only in the areas where the excess electrons are located but also in regions between the excess electrons and the interfacial Sn atoms. These continuous electron densities, spanning from the Sn atoms to the sites of localized excess electrons, further suggest a covalent-like feature arising from the hybridization of excess electrons and Sn states. While bilayer PbO exhibits an ELF distribution similar to that of SnO (Fig. S1), its interlayer electron localization is less pronounced, most likely, due to a smaller interlayer distance. In the following paragraphs, we discuss the formation mechanism and properties of these excess electrons and their relation to the electronic bandstructures.

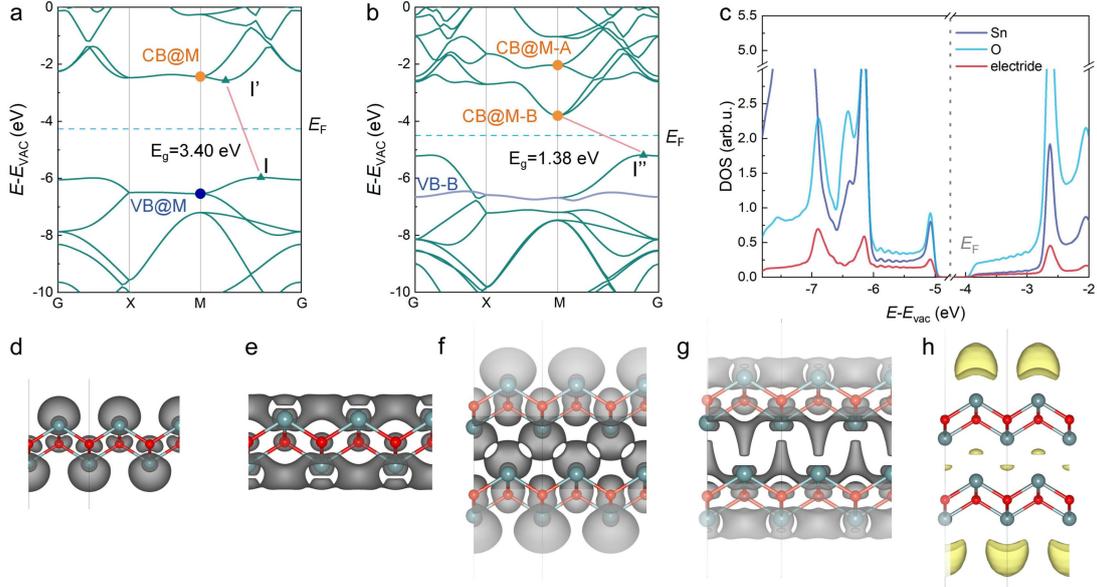

**Figure 2. Electronic structures of monolayer and bilayer SnO.** (a) Bandstructure of monolayer SnO. The vacuum energy level is set to zero, and the Fermi level is marked with a blue dashed line. The green triangles and solid pink lines indicate the VBM and CBM, with the bandgap value represented as $E_g$. The CB and VB at the M-point are indicated by orange and blue circles. (b) The same scheme of plots as that of panels (a) for the bandstructure of bilayer SnO. (c) Total and projected density of states (DOS) of SnO and the excess electrons. (d-e) Visualized wavefunction norms of the VB (d) and CB (e), marked with blue and orange circles in (a), respectively. The isosurface contours are $8×10^{-3}$ e/Bohr$^3$. (f) Partial charge density of the VB in the energy range from -6.5 to -7 eV, denoted as VB-B. (g) Visualized wave function norms for the bilayer SnO at the CBM (BL-CB@M-A)..The isosurface contours are $2×10^{-3}$ e/Bohr$^3$. (h) Differential ELF before and after doping. The isosurface contours are $5×10^{-5}$.

Monolayer SnO retains semiconducting properties observed in its bulk form, as indicated by its electronic bandstructure shown in Fig. 2a. The valence band maximum (VBM) sits at I (0.2, 0.2, 0), while the conduction band minimum (CBM) lies at I' (0.375, 0.375, 0), indicating an indirect band gap of 3.40 eV. Along path X-M-I (I'), both VB and CB, especially CB, exhibit minimal dispersions, implying low mobility for carriers flowing through these bands. Upon stacking two layers in to a bilayer, as depicted in Fig. 2b, significant band splitting occurs, particularly at the VBM and CBM with an indirect band gap of 1.38 eV. The VBM shows an approximately 0.5 eV splitting and the CBM exhibits an around 1.8 eV splitting at the M-point. Additionally, the VBM shifts to I''(0.085, 0.085, 0) and the CBM is now located at the M point. Such large energy splitting in the eV level indicates exceptionally strong interfacial electronic couplings and the parabolic shape of the CB appears highly comparable to that of $Mg_2Si$, a prototypic potential electride[36]. To further explore the interfacial coupling, empty spheres were placed at positions of excess electrons for density of states projection. As shown in Fig. 2c, the DOSs projected onto interfacial Sn atoms and electride spheres indicate clear electronic hybridization between the excess electrons and surrounding interfacial Sn states for either occupied or unoccupied bands near the Fermi level. This hybridization, again, verifies the SnO is, at least, a hybrid electride.

Figure 2d and 2e illustrates the wavefunction norms of the VB (Fig. 2d) and CB (Fig. 2e) at the M-point of the SnO monolayer, each of which is 2-fold degenerated. The CB states (ML-CB@M, Fig. 2e) are primarily comprised of Sn $5p_x$ and $5p_y$ orbitals, forming a delocalized tetragonal network in each Sn layer (Fig. S2a), while the VB states (ML-VB@M, Fig. 2d) are hybridized by the Sn $5s$ and $5p_z$ orbitals, along with the O $2p_x$ and $2p_y$ orbitals (Figs. S2a and S2b). In the bilayer, the ML-VB@M (ML-CB@M) states from the two layers are electronically hybridized at the vdW interface, forming interlayer anti-bonding and bonding states, denoted BL-VB@M-A and BL-VB@M-B (BL-CB@M-A and BL-CB@M-B). Figure 2f depicts the wave function norm square of states integrated from -6.5 to -7.0 eV, reflecting the feature of BL-VB-B. The plot convincingly demonstrates significant wavefunction overlaps between the interfacial excess electron sites and the Sn metal atoms, which is a smoking-gun piece of evidence for the covalency in the SnO electride.

Besides the occupied states, the wave function norm square of the BL-CB@M-B states (Fig. 2g), showing a parabolic feature in the BL bandstructure (Fig. 2b), indicates an unoccupied state confined within the vdW interface. The confined state develops from the delocalized in-plane $p_{x/y}$ state (shown in Fig. 2d) of one layer towards the Sn atoms of the other layer, identifying it as a potential electride state[22]. As discussed in the literature, electrides can be categorized into conventional electrides and potential electrides, characterized by whether the confined localized state is occupied (conventional electrides) or unoccupied (potential electrides). To further confirm the existence of potential electride behavior, we doped the bilayer SnO with an additional electron, which occupied a portion of the conduction band. We then calculated the ELF for both the doped and undoped bilayers, as shown in Supplementary Figure S5. Figure 2h illustrates the differential ELF before and after doping. It is evident that after doping with one electron, the charge localization on the outer Sn atoms decreases, while the excess electrons between layers become more localized. This indicates that the doped electron tends to accumulate in the interlayer region, further strengthening the potential electride characteristics. The coexistence of these two types of electrides has not been observed in the literature, however, they coexist in bilayer SnO. We thus term it a "bipolar electride", reflecting the electride states are presented in both the VB and CB.

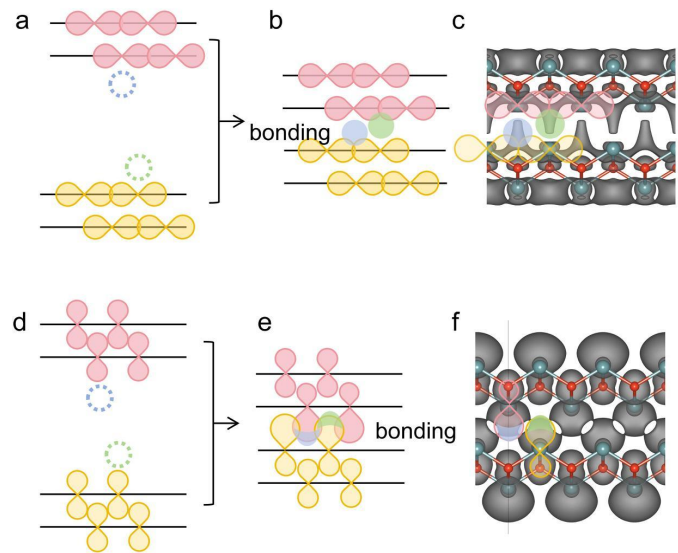

**Figure 3. Mechanism of hybridization for the systems.** (a-b) Bonding states of $p_x$&$p_y$ orbitals of Sn atoms in bilayer case. Pink and orange dumbbells represent the orbitals of Sn atoms in the upper and lower layers, respectively. The blue and green circles represent the orbital positions of the excess electrons. (c) Visualized wave function norms overlaid with specific orbital models. (d-f) Bonding states of $p_z$ orbitals of Sn atoms, visualized using the same plotting scheme.

We next discuss the formation mechanism of bilayer SnO, a semiconducting electride with covalent hybridization and bi-polar characteristics. In monolayer SnO, two of the four valence electrons from each Sn atom are transferred to O atoms, leaving the in-plane Sn $5p_x$ and $5p_y$ orbitals as 2-fold degenerate empty states (see Fig. S3). These orbitals undergo in-plane hybridization, forming a covalent-like in-plane delocalized state (ML-CB@M), which is primarily distributed in the interstitial regions between Sn atoms, as illustrated in Fig. 3a. The remaining two valence electrons from Sn, together with two electrons contributed by O, occupy two degenerated states (ML-VB@M) primarily formed by vertical hybridization of Sn $5p_z$ and $5s$ orbitals (see Fig. 3b for hybridization and Fig. S2 for orbital projection). The distinct difference in hybridization introduces strong in-plane and out-of-plane anisotropy to the ML-VB@M and ML-CB@M states, resulting in significant energy splitting and serving as the primary origin of the semiconducting behavior.

Upon stacking the two layers to form a bilayer, the in-plane delocalized ML-CB@M states form side-by-side covalent-like overlaps in the interlayer region. The wavefunction norm square extends from the interstitial sites of Sn in one layer toward Sn atoms in the other layer, as shown by the blue and green shaded regions in Fig. 3b). This results in a cone-shaped wavefunction norm square of the hybridized bonding state, BL-CB@M-B, confined at the 2c position within the interlayer region, as depicted in Fig. 3c. Similarly, the out-of-plane ML-VB@M states from both layers covalently hybridize in a side-by-side manner within the bilayer. In this case, the Sn $5p_z$ orbitals extend into the interlayer region, occupying the 2c position and overlapping laterally. This overlap is illustrated by the blue and green shaded regions in Fig. 3e-3f, forming in a quasi-2D distribution of excess electron primarily contributed by the BL-VB@M-B state. These structural and electronic characteristics explain why SnO exhibits a novel bi-polar nature as an electride, with covalent hybridization occurring between the excess electrons and the metal Sn states.

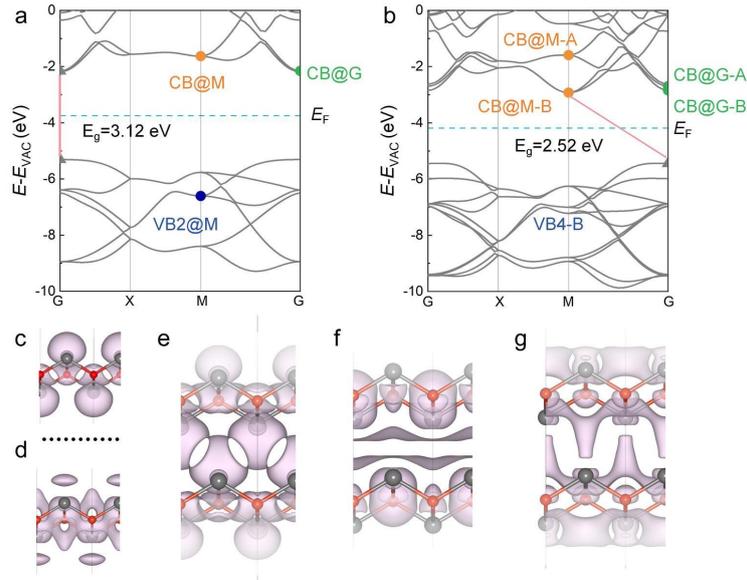

**Figure 4. Electronic structures of PbO.** (a-b) Bandstructure of monolayer (a) and bilayer (b) PbO, with labels identical to those in Figure 2. (c-d) Visualized wavefunction norms for the bands marked in (a). (c) shows a valence band at the M-point, marked with a blue circle, while (d) displays the CBM at the Γ-point, indicated by a green circle. The isosurface contours are 0.002 e/Bohr$^3$. (e-g) Visualized wavefunction norms for the splitting VB at the M-point (e) and CB at the G-point (f) and M-point (g), showing the bonding states. The isosurface contours are 0.003 e/Bohr$^3$.

The PbO mono- and bi-layers exhibit comparable but slightly different properties to those of SnO. Figure 4a shows the bandstructure of monolayer PbO, showing a direct bandgap at the G point, with flat bands visible in CB along the X-M-I path, although the overall shape of those plotted bands is comparable to that of monolayer SnO. The largest difference between the bandstructure of PbO and SnO lies in the VBM and CBM of PbO reside at the G point. Upon forming a bilayer, significant band splitting occurs similarly to that in SnO, transforming the direct bandgap (3.12 eV) into an indirect one (2.52 eV) (Figure 4b). Another difference is related to the band alignment that the ML-VB@M states at the M point is a state comprised by O 2p orbitals. However, the ML-VB2@M states at the M point (marked by the blue box in Fig. 4a) are comprised of the Pb 6$p_z$ and 6$s$ orbitals (Fig. 4c), showing a similar shape to that of ML-VB of SnO. The ML-CB state at the M point (ML-CB@M) is composed of Pb 6$p_x$ and 6$p_y$ states, consistent with that of SnO. However, the ML-CB state at the G point (ML-CB@G), the CBM, is composed of another state visualized in Fig. 4d. These ML states hybridize across interlayer, form bonding (Fig. 4e-4g) and antibonding states (Fig. S5c-e). The BL-VB4-B (bonding state of two ML-VB2) and the BL-CB@M-B state show wavefunction overlaps at the 2c position within the interlayer regions, similar to those of SnO. An interesting new bonding state, marked BL-CB@G-B (Fig. 4f), shows a highly extended state confined at the interlayer region. This state also contributes to the wavefunction norm square at the 2c positions, indicating the presence of multiple unoccupied conduction electride states in PbO.

| str | lattice (Å) | band gap (eV) | VBM | | | | CBM | | | |
|---|---|---|---|---|---|---|---|---|---|---|
| | | | coordinates | $M_e$ ($m_0$) | $E_l$ (eV) | Mobility ($cm^2/V \cdot s$) | coordinates | $M_e$ ($m_0$) | $E_l$ (eV) | Mobility ($cm^2/V \cdot s$) |
| SnO | 3.796 | 1.38 | I (0.85,0.85,0) | 1.11 | 0.27 | 3270 | M (0.5,0.5,0) | 0.57 | 0.78 | 1470 |
| PbO | 3.986 | 2.52 | I (0.1,0,0) | 4.61 | 0.44 | 15 | M (0.5,0.5,0) | 1.66 | 1.02 | 24 |

Table 1. Calculated properties of bilayer SnO and PbO. The lattice constants, bandgap values, as well as the specific characteristics of the VBM and CBM are listed. $M_e$ is the effective mass, $E_l$ represents the deformation potential, and "Mobility" refers to the carrier mobility.

The potential electride states in either SnO or PbO introduce highly dispersive states to the CBs, e.g. 0.57 $m_0$ for bilayer SnO, which suggest high mobility of electrons. We thus predicted their longitudinal acoustic phonon limited carrier mobility for both electron and hole in SnO and PbO bi-layers. Among these predicted results, bilayer SnO exhibits high carrier mobility for either hole (3270 $cm^2$/Vs) or electron (1470 $cm^2$/Vs), which are two orders magnitude larger than those for bilayer PbO. These high carrier mobilities are primarily due to small deformation potentials $E_l$ of 0.27 and 0.78 eV, and small effective mass of 1.11 and 0.57 $m_0$ for the VBM and CBM.

**Discussion and Conclusions**

As reported in the previous study on 2D semiconducting electrides, increased metal electronegativity reduces its ionicity and enhances its overlap with excess electrons, leading to the formation of semiconducting electrides. The Pauling electronegativities of Sn and Pb are 1.96 and 2.33, both higher than previously examined high electronegativity metals such as Sc (1.36), Y (1.22), and Al (1.61). Consistently, the covalency in the SnO and PbO electrides is found stronger than that in $Sc_2C$, $Y_2C$, or $Al_2C$. The excess electrons of SnO or PbO covalently hybridize with Sn or Pb states, exhibiting a maximum bandgap of up to 2.52 eV in PbO (1.38 eV for SnO). Therefore, such p-block metals exhibit unique advantages in forming electrides, which enhances strong hybridization between excess electrons and metal cation states, manifesting as covalent-like rather than typically ionic interactions. A recent high-throughput computational screening identified only 13 monolayer semiconductors with mobilities exceeding 1400 $cm^2/Vs$[37]. In our study, unlike monolayer SnO which exhibits ferroelasticity, bilayer SnO demonstrates high mobilities over 1400 $cm^2/Vs$ for both electrons (1470 $cm^2/Vs$) and holes (3270 $cm^2/Vs$). This "bipolar" electride, characterized by its high ambipolar mobility, offers advantages over most of those 13 high mobility semiconductors, which are limited to only one type of charge carrier.

In summary, we have identified and characterized a class of bi-polar semiconducting electrides in p-block metal oxides bilayers, particularly SnO and PbO. They exhibit highly localized excess electrons within the interlayer regions, as confirmed by our ELF analysis. Unlike conventional electrides, where confined excess electrons interact electrostatically with surrounding cations, the interlayer excess electrons in SnO and PbO engage in covalent-like interactions with neighboring metal cations. Such interactions induce semiconducting properties in SnO and PbO, while discovery of the covalency in the electron-cation interactions extends the family of the hybrid electrides. In addition to the hybrid electride property, they are also potential electrides, exhibiting an unoccupied free-electron-like state confined within their interlayer regions. We proposed the concept of "bipolar electrides" to describe the coexistence of hybrid and potential electride properties. The calculated effective masses and mobilities, especially for the conduction bands where the FEL states are present, emphasize the high potential of SnO and PbO for efficient charge transport. Our findings expand the knowledge on the electronic behavior of 2D semiconducting electrides, paving the way for future studies on p-block metal electrides and their integration into advanced electronic devices.

**Methods**

Our density functional theory (DFT) calculations were performed using a meta-generalized generalized gradient approximation (meta-GGA) for the exchange correlation potential in the form of r2SCAN[38] (regularized-restored SCAN), the projector augmented wave method[44], and a plane-wave basis set as implemented in the Vienna ab-initio simulation package (VASP)[45]. The energy cutoff for the plane-wave basis-sets was set to 700 eV for variable volume structural relaxation and electronic structure calculations. All atoms, lattice volumes, and shapes were allowed to relax until the residual force per atom was below 0.001 eV/Å. A vacuum layer exceeding 15 Å in thickness was employed to reduce imaging interactions between adjacent supercells. A M-centered $k$-mesh of 14×14×12 and 14×14×1 was used to sample the first Brillouin zone of the unit cell for bulk and layered structures, respectively. The Gaussian smearing method with a $\sigma$ value of 0.01 eV was applied for all calculations except for the DOS calculation, which utilized a $\sigma$ value of 0.03 eV. DCD was calculated using the formula $\Delta\rho_{DCD} = \rho_{bilayer} - \rho_{upper\ layer} - \rho_{lower\ layer}$. Here $\rho_{bilayer}$ represents the total charge density of the bilayer SnO or PbO, while $\rho_{upper\ layer}$ and $\rho_{lower\ layer}$ correspond to the total charge densities of the individual upper and lower monolayers, respectively. The differential ELF was calculated using the formula $\Delta ELF = ELF_{before\ doping} - ELF_{after\ doping}$. For the calculation of effective masses, specific high-symmetry paths in the Brillouin zone were selected for fitting. For bilayer SnO, the VBM was fitted along the path from (0.25, 0.25, 0) to the Γ-point, while the CBM was fitted along the path from (0.5, 0.325, 0) to (0.455, 0.455, 0). In the case of bilayer PbO, the VBM was fitted along the path from (0.02, 0, 0) to (0.155, 0, 0), and the CBM was fitted along the path from (0.5, 0.335, 0) to (0.4, 0.4, 0). These paths were chosen based on the curvature of the bands near their extrema to ensure accurate extraction of effective masses.

We considered the scattering due to longitudinal acoustic (LA) phonons, using the following equation:

$$\mu_{2D} = \frac{\pi e \hbar^4 C_{2d}}{\sqrt{2}(k_B T)^{3/2}(m^*)^{5/2}(D_A)^2} F$$

In this equation, $C_{2d}$ is the elastic constant, valued at 13.93 J/m$^2$, m$^*$ represents the effective mass; $E_l$ is the deformation potential; and $F$ denotes a crossover function[39-41].

**Acknowledgements**

We gratefully acknowledge the financial support from the Ministry of Science and Technology (MOST) of China (Grant No. 2023YFA1406500), the National Natural Science Foundation of China (Grants No. 11974422 and 12104504), the Fundamental Research Funds for the Central Universities, and the Research Funds of Renmin University of China [Grants No. 22XNKJ30] (W.J.). J.D. was supported by the Outstanding Innovative Talents Cultivation Funded Programs 2023 of Renmin University of China. All calculations for this study were performed at the Physics Lab of High-Performance Computing (PLHPC) and the Public Computing Cloud (PCC) of Renmin University of China.